\documentclass[twocolumn,showpacs,amsmath,amssymb]{revtex4}

\usepackage{graphicx}
\usepackage{dcolumn}
\usepackage{bm}

\begin{document}

\preprint{APS/123-QED}

\title{
Spiral charge frustration in 
molecular conductor (DI-DCNQI)$_2$Ag
}

\author{
Hitoshi Seo$^{1,*}$ and Yukitoshi Motome$^2$
}
\affiliation{%
$^{1}$Synchrotron Radiation Research Center, Japan Atomic Energy Agency,
SPring-8, Hyogo 679-5148, Japan\\
$^{2}$Department of Applied Physics, University of Tokyo, Bunkyo-ku, Tokyo 113-8656, Japan
}%

\date{\today}

\begin{abstract}
We theoretically study the effect of spiral-type 
charge frustration 
in a quasi-one-dimensional molecular conductor (DI-DCNQI)$_2$Ag. 
We clarify how the spiral frustration in the interchain Coulomb repulsion is relieved
and leads to a self-organization of complex charge-lattice ordered chains, 
in agreement with the recent 
synchrotron X-ray study
[T. Kakiuchi {\it et al.}, Phys. Rev. Lett. {\bf 98}, 066402 (2007)]. 
In addition, we find that 
a keen competition between charge and lattice degrees of freedom under the frustration
gives rise to a characteristic temperature within the ordered 
phase, below 
which a drastic 
growth of molecular displacements occurs. 
Our results 
enlighten the relevance of 
the spiral frustration and 
provide 
a possible reconciliation
among puzzling experimental data. 
\end{abstract}

\maketitle


Geometrical frustration 
in charge degree of freedom, i.e., charge frustration, 
has been a fundamental problem
in condensed matter physics, 
since Anderson argued its importance 
in the 
Verwey transition in magnetite~\cite{Anderson1956PR}.
Recently, it is attracting renewed interests
partly because 
exotic behaviors are found in several molecular conductors, 
where frustrated lattice structures are often formed. 
A typical example is the bis(ethylenedithio)tetrathiafulvalene 
(BEDT-TTF) based compounds,
in which strong frustration in the two-dimensional triangular lattice
makes the charge ordering (CO) unstable~\cite{Mori1998PRB} and
intriguing transport phenomena are 
observed 
in the critical region
\cite{Sawano2005Nature,Yamaguchi2006PRL}.

Quasi-one-dimensional (Q1D) molecular conductors, however, 
have been less examined from the viewpoint of 
frustration; it 
is usually supposed that the 1D chains are well 
decoupled, 
offereing canonical systems to study 
1D physics such as Tomonaga-Luttinger liquids, 
charge and spin density-waves, 
and so on
~\cite{Gruner1994}.
Nevertheless, 
the Q1D compounds 
also 
often retain 
a ``hidden'' three-dimensional (3D) 
frustration in packing the 1D chains. 
Such interchain frustration effect has been 
studied within extended Hubbard-type models 
on frustrated zigzag ladders~\cite{Seo2001PRB,Nishimoto2003PRB,Clay2005PRL}. 

We consider herea typical Q1D molecular conductor (DI-DCNQI)$_2$Ag~\cite{Hiraki1996PRB},
which provides a different avenue to interchain frustration. 
This compound consists of 
1D chains of DI-DCNQI (2,5-diiodo-$N,N'$-dicyanoquinonediimine) molecules 
whose electronic structure near the Fermi level is described 
by a Q1D band at quarter filling~\cite{Miyazaki1995PRL}. 
A prominent feature of this system is the phase transition at $T_{\rm c} \simeq 220$~K.
It was observed that the $^{13}$C-NMR line splits below $T_{\rm c}$,
which was ascribed to a Wigner-crystal type CO 
with a twofold period along the chains~\cite{Hiraki1998PRL}.
Although the simple CO scenario was 
supported by
theoretical studies in 1D models~\cite{Seo2006JPSJreview}, 
there have been many puzzling behaviors found in experiments.

One is the
lattice displacements suggested by
optical spectra analyses~\cite{Meneghetti2001SM,Yamamoto2005PRB}: 
The IR and Raman 
spectra are complicated and not well fitted by such simple CO
without assuming any lattice distortions.
Another 
unaccountable 
behavior is seen in the NMR line shape~\cite{Hiraki1998PRL}.
There appears an extraordinarily wide broadening
below $\sim$100~K
in the outer line
with a larger shift, 
which can hardly be ascribed to critical fluctuations
related to the magnetic ordering at 
a considerably lower temperature ($T$) of 
5~K. 
Furthermore, a small peak starts to 
grow at around zero shift 
in a similar 
$T$ range. 
Such a characteristic $T$ range within 
the ordered phase 
was 
also observed in transport properties.
The $1/T$ derivative of the log of resistivity 
shows, in addition to a peak structure at $T_{\rm c}$, 
a broad hump centered 
at around 100~K~\cite{Itoh2004PRL}.
In the same region, the dielectric permittivity exhibits a peak and
large frequency dependence~\cite{Nad2004JPCM}.
All these behaviors, 
implying intrinsic changes 
at a much lower $T$ than $T_{\rm c}$, 
are difficult to understand from the simple 
1D CO picture. 

\begin{figure}
 \centerline{\includegraphics[width=8.4truecm]{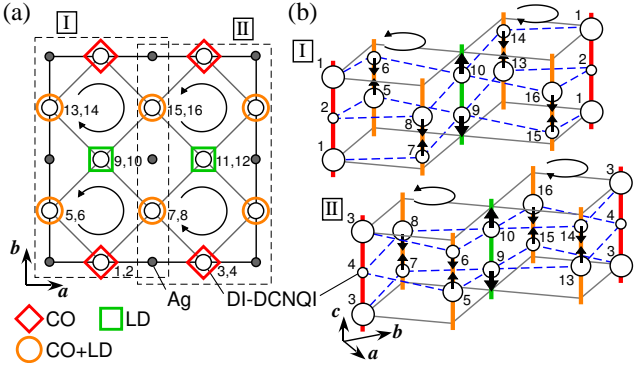}}
\caption{(color)
The 
charge-lattice ordered state (the mixed state) 
 in (DI-DCNQI)$_2$Ag~\cite{Kakiuchi2007PRL}. 
(a)~DI-DCNQI and Ag sites 
viewed from 
the 
chain direction 
($c$ axis),
where the arrangement of chains with different 
orders
are indicated; 
 CO and LD represent charge order and lattice dimerization, respectively. 
(b)~Side view 
of 
the DI-DCNQI sites 
 where 
the size of circles and arrows 
represent the charge density and the lattice displacement, respectively. 
The blue dotted 
lines in (b) are the spiral $V'$ bonds in eq.~(\ref{eq1}). 
}
\label{fig1}
 \vspace*{-1em}
\end{figure}

A clue to a possible origin of these features 
was found by 
a recent experimental study based on synchrotron X-ray 
crystal structure analysis at 50~K
\cite{Kakiuchi2007PRL}.
It was claimed that 
the ordered structure 
is not a simple CO:
Instead it is 
a periodic array of DI-DCNQI chains 
with mixed 
patterns of charge-lattice symmetry breaking
as shown in Fig.~\ref{fig1}, 
which 
we call 
the ``mixed state.''
The importance of 
frustration in the lattice structure was pointed out; 
if all the chains exhibit the simple twofold CO, 
they would suffer from a spiral-type charge frustration among neighboring chains, 
since the DI-DCNQI molecules 
are displaced
with each other by 1/4 of lattice spacing along the chain 
as shown in Fig.~\ref{fig1}(b) (see also Fig.~4 of ref.~\cite{Kakiuchi2007PRL}). 
However, 
it is not 
clear how the mixed state is stabilized and, 
furthurmore, 
what causes the peculiar $T$ dependences in the ordered phase mentioned above.

Theoretically, 
interplay 
between charge and lattice degrees of freedom
in Q1D molecular conductors 
has been studied on the basis of 
1D or Q1D extended Hubbard models at quarter-filling~\cite{Seo2006JPSJreview}.
In particular, 
recent studies 
showed 
that CO and lattice dimerization (LD) compete with each other 
but they coexist in some cases
\cite{Seo2007JPSJ}. 
These three states, CO, LD, and their coexistence, are in fact realized 
but in different chains in the mixed 
state shown in Fig.~\ref{fig1}. 
The spiral frustration 
is expected to be a key factor for the nontrivial mixture of 
 orderings 
 out of uniform and equivalent chains, 
however, theoretical studies 
taking account of it have not been done so far.

In this study, we investigate 
the effect of the interchain spiral frustration in Q1D systems
for a model 
 of (DI-DCNQI)$_2$Ag. 
We clarify how 
the mixed state in Fig.~\ref{fig1}
is stabilized under the charge frustration.
We also calculate finite-$T$ properties of the charge-lattice coupled phenomena
and propose a possible 
scenario to reconcile 
the puzzling experimental data in this compound.

We consider a Q1D model which takes account of
 the 3D interchain spiral structure in the 
actual compound.
Each 1D chain is represented by 
the extended Peierls-Hubbard model
 at quarter filling~\cite{Seo2006JPSJreview}. 
Here, in order to concentrate on the competition 
between charge and lattice degrees of freedom, 
 we take
the limit of strong onsite Coulomb interaction 
 and leave out the spin degree of freedom~\cite{Ovchinnikov1973Sov}. 
Namely, 
 we study a half-filled spinless fermion model 
 coupled to the lattice, 
whose Hamiltonian reads
\begin{align}
& {\cal H} = \sum_{\langle i,j \rangle_{\rm 1D}} 
  \Big\{ 
  t_{ij} 
  \big( a^\dagger_{i} a_{j}^{} + \mathrm{h.c.} \big) 
  + 
  V_{ij} 
  \, n_{i} n_{j} \Big\} \nonumber \\ 
& \hspace{5mm} + \sum_{\langle i,j \rangle_\perp'}  
V'_{ij} 
\, n_{i} n_{j} 
  + \sum_{\langle i,j \rangle_\perp''}  
  V''_{ij} 
  \, n_{i} n_{j}
  + \frac{K}{2} \sum_{i} u_{i}^2,
\label{eq1}
\end{align}
 where $a^\dagger_{i}$ ($a_{i}$) is the creation (annihilation) operator 
 for a spinless fermion 
 at the $i$th DI-DCNQI site, and $n_i=a^\dagger_{i}a_{i}$. 
Here, the first term is the 1D part where the sum on $\langle i,j \rangle_{\rm 1D}$ 
 is taken for nearest-neighbor sites along the chains. 
The next two terms are the interchain Coulomb interactions 
 where $\langle i,j \rangle_\perp'$ are the nearest %
interchain site-pairs 
 shown by the blue dotted lines in Fig.~\ref{fig1}(b), 
 and $\langle i,j \rangle_\perp''$ are the next-nearest 
interchain
pairs 
 such as sites 1-8, 2-7, etc., 
 while interactions of farther range are neglected for simplicity.  
Note that both terms 
undergo the spiral frustration.
The transfer integrals and the Coulomb repulsions are coupled 
with the lattice displacement 
$u_i$ of each molecule, 
  as 
 $t_{ij}=t(1+\alpha \, \delta u_{ij})$, 
 $V_{ij}=V(1+\beta \, \delta u_{ij})$, 
 $V'_{ij}=V'(1+\beta' \, \delta u_{ij})$, and 
 $V''_{ij}=V''(1+\beta'' \, \delta u_{ij})$, 
where $\delta u_{ij} = u_i - u_j$~\cite{note0}. 
 The last term in eq.~(\ref{eq1}) is the elastic energy, and
 the displacements $u_i$ are treated as classical variables within 
the 
 adiabatic approximation.
In the following, 
 we restrict ourselves to $u_i$ being uniaxial, i.e., 
 the molecules are allowed to move only in the chain direction, 
by considering
the experimental results 
\cite{Kakiuchi2007PRL}. 

We treat the Coulomb interaction 
within Hartree-Fock approximation. 
The size of unit cell we consider is 
 the experimental one which contains 8 chains with 2 sites each, 
 namely, 16 sites as shown in Fig.~\ref{fig1}. 
The order parameters, $\langle n_i \rangle$, $\langle a_i^\dagger a_j \rangle$,
and $u_i$ are self-consistently determined with the precision of 10$^{-6}$.
We
set $t=1$ and $\alpha=1$ (absorbed in the notation of $u_i$), 
 and fix 
 $V=1.5$, $V'=0.75$, $V''=0.15$, $\beta=0.5$, $\beta'=0.0325$, 
 and $\beta''=0.0975$~\cite{note}, 
 while varying $K^{-1}$, which represents the softness of the lattice. 
We will comment on the choice for 
the intersite Coulomb repulsions  
later. 
\begin{figure}
 \centerline{\includegraphics[width=8.4truecm]{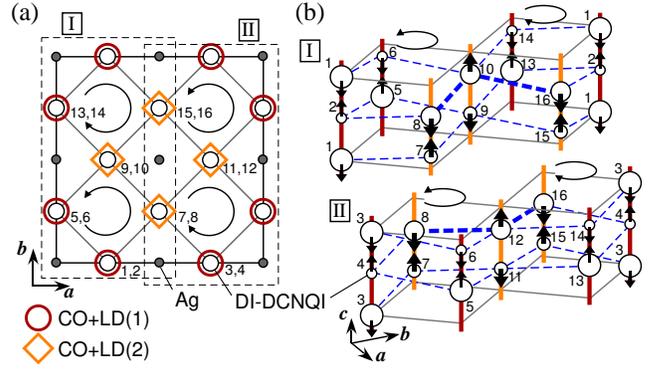}}
\caption{(color) 
The CO+LD state. 
Notations are the same as in Fig.~\ref{fig1}, 
whereas CO+LD(1) and CO+LD(2) in (a) indicate the two kinds of chains, 
 both with charge order as well as lattice dimerization but with different degrees. 
The thick $V'$ bonds in (b) 
suffer from frustration in Coulomb repulsion 
between charge rich sites. 
}
\label{fig2}
 \vspace*{-1em}
\end{figure}%

\begin{figure}
 \centerline{\includegraphics[width=8.7truecm]{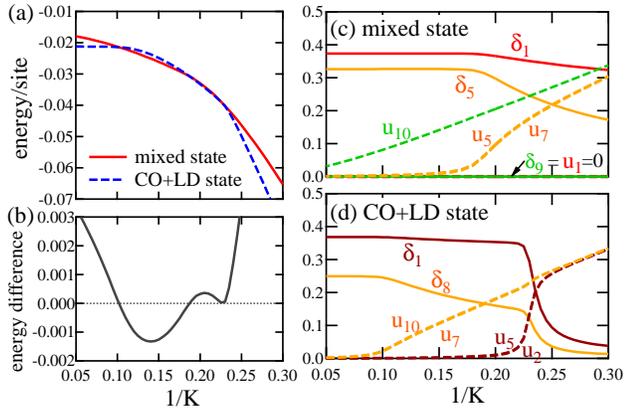}}
\vspace*{-1.8em}
\caption{(color)
The ground state properties of 
the mixed state and the CO+LD state. 
$K^{-1}$ dependences of (a) the ground state energies, 
(b) energy difference between the two states, 
 and (c), (d) charge disproportionations 
 $\delta_i = \langle n_{i} \rangle -1/2$
 and lattice displacements $u_i$ 
 for representative sites in the two solutions. 
$\delta_i$ and $u_i$ for other sites are equal or have opposite sign 
 with equal absolute value 
 to the plotted ones 
 (see Figs.~\ref{fig1} and \ref{fig2}). 
Note that $(u_5, u_7)$ in (c) and $(u_2, u_5)$ and $(u_7, u_{10})$ in (d)
take almost the same but not identical values 
[see also the insets of Figs.~\ref{fig4}(c) and (d)]. 
}
\label{fig3}
 \vspace*{-1em}
\end{figure}

First let us discuss the results 
for 
the ground state 
properties. 
We find that 
 the charge-lattice ordering equivalent to 
the mixed state pattern in Fig.~\ref{fig1} is 
indeed 
 stabilized 
 over others 
 in a certain range of $K^{-1}$, 
among the 
several 
solutions 
self-consistently obtained. 
Otherwise, another solution gives the lowest energy: 
In this state, 
all chains show both CO and LD
as shown in Fig.~\ref{fig2},
which we call here the CO+LD state. 
Both of these states are insulating. 
Figure~\ref{fig3} shows
the ground state energies as well as 
 the charge disproportionations
$\delta_i = \langle n_{i} \rangle - 1/2$
 and the lattice displacements
$u_i$ for these two solutions as a function of $K^{-1}$. 
For $K^{-1} < 0.10$, 
 the CO+LD state is the ground state: 
 As shown in Fig.~\ref{fig3}(d),
 the lattice displacements $u_i$ are 
almost negligibly small 
in this region; 
 therefore, it is essentially regarded as 
 a bundle of 1D CO states stabilized by the intrachain Coulomb 
repulsion $V$. 
This state suffers from the charge frustration 
 in the thick 
$V'$ bonds in Fig.~\ref{fig2}(b). 
On the other hand, for $K^{-1} > 0.19$, 
 the CO+LD state is the ground state as well~\cite{note2}; 
 when $K^{-1}$ is large, 
 the charge disproportionations $\delta_i$ become small, 
 and hence, 
 it is essentially a LD state 
 stabilized by the intrachain charge-lattice coupling. 
In between these two,
 the mixed state gives a lower energy for $0.10 < K^{-1} < 0.19$. 
This state emerges in the transient regime from the 
CO-dominant state for small $K^{-1}$ to the LD-dominant state for large $K^{-1}$,
both of which undergo frustration.
There, 
the system finds
a way to relieve the frustration 
 by 
self-organizing arrays of 
the CO, LD, and 
coexisting chains. 

\begin{figure}
 \centerline{\includegraphics[width=8.7truecm]{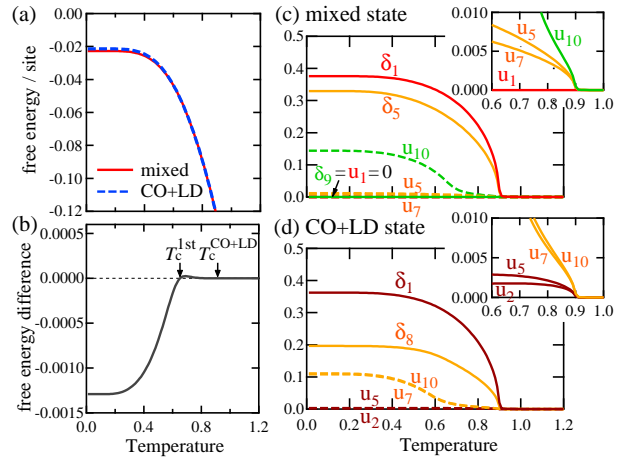}}
\vspace*{-1.8em}
\caption{(color)
Temperature dependences of (a) the free energies, 
(b) free energy difference, 
  and (c), (d) charge disproportionations 
  $\delta_i = \langle n_{i} \rangle -1/2$ 
 and lattice displacements $u_i$ for the two solutions 
 for $K^{-1}=0.15$. 
Insets show enlarged view for $u_i$ near the transition temperatures.}
\label{fig4}
 \vspace*{-1em}
\end{figure}

\begin{figure}[b]
 \centerline{\includegraphics[width=7.5truecm]{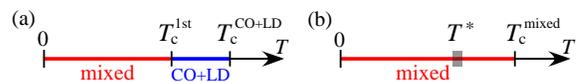}}
\caption{(color)
Schematic 
illustration 
of the proposed two scenarios 
for the phase transition(s) 
 in (DI-DCNQI)$_2$Ag.}
\label{fig5}
\end{figure}

Next we show how these 
coexistences and 
competitions evolve at finite $T$. 
In Fig.~\ref{fig4}(a), 
 $T$ dependences of the free energies for 
the 
two solutions 
 are shown at $K^{-1}=0.15$, 
for which 
the ground state is the mixed state.
As increasing $T$,
 the free energy difference between the two states 
 becomes smaller, 
and at $T_{\rm c}^{\rm 1st} = 0.66$, a first-order transition takes place 
above which 
the CO+LD state gives lower free energy [see also Fig.~\ref{fig4}(b)]. 
Finally, 
 for $T > T_{\rm c}^{\rm CO+LD} = 0.91$ the system turns into the uniform metallic state.
The sequence of the phase transitions is summarized in Fig.~\ref{fig5}(a). 

An important point is that the free energy difference is extremely small
for $T_{\rm c}^{\rm 1st} <T <T_{\rm c}^{\rm CO+LD}$, 
much less than 10$^{-4}$ as shown in Fig.~\ref{fig4}(b). 
Another key observation is that
$T$ dependences of $\delta_i$ and $u_i$ 
 are similar between these two solutions
 as plotted in Figs.~\ref{fig4}(c) and (d).
Their transition temperatures 
from the high-$T$ uniform phase
are almost the same
($T \simeq 0.91$), 
 and by decreasing $T$, charge disproportionations $\delta_i$ 
 develop first
 while lattice distortions $u_i$ remain 
 suppressed.  
 When $T$ is decreased down to 
 $T^* \simeq 0.7$, 
then the 
$u_i$ rapidly develop. 
Note that this crossover temperature $T^*$ almost coincides with 
$T_{\rm c}^{\rm 1st}$.
The first order phase transition as well as 
the finite-$T$ features described above 
are always seen in our calculations 
when the mixed state is the ground state. 
From these results at finite $T$,
we propose two different scenarios for 
what is taking place in the actual 
compound (DI-DCNQI)$_2$Ag.
One is 
that, as in our calculation, 
the system exhibits two phase transitions.
The first one is the transition at $T_{\rm c}=220$~K,
which corresponds to $T = T_{\rm c}^{\rm CO+LD}$,
and
the second occurs at $T_{\rm c}^{\rm 1st}$
well below $T_{\rm c}$ 
as indicated in Fig.~\ref{fig5}(a); 
yet this latter transition 
has not been observed in experiments so far.
The other scenario is that the CO+LD state 
in our calculation 
is destabilized 
if fluctuations are 
taken into account
beyond the mean-field level, and 
the mixed state gives a lower free energy in the entire $T$ range 
within the ordered phase.  
Then, we predict that there is 
only one transition to the mixed state; 
$T_{\rm c}$ = 220~K corresponds to $T_{\rm c}^{\rm mixed}$,
which is followed by 
a crossover behavior at $T^*$, 
instead of the successive transitions, 
as illustrated in Fig.~\ref{fig5}(b). 
This latter scenario is based on the observation that 
CO suffering from frustration is often destabilized
by fluctuations~\cite{Merino2005PRB,Seo2006JPSJreview}, 
and 
the free energy difference is already very tiny in the present mean-field results.

In both 
of these 
scenarios, an additional characteristic temperature 
is predicted: 
$T_{\rm c}^{\rm 1st}$ in the former and $T^*$ in the latter.
In either case, the system undergoes drastic changes 
there 
in the lattice sector, 
i.e., 
a rapid development of the lattice displacements 
within the ordered phase. 
This is suggestive in considering 
the puzzling
experimental results 
mentioned above.
The rapid growth and 
the expected 
large fluctuations in the lattice 
displacements 
may explain a peak and frequency dependence of 
the dielectric permittivity~\cite{Nad2004JPCM} as well as
the anomalous broadening and an additional peak
in the NMR spectra~\cite{Hiraki1998PRL}.
A hump in the $1/T$ derivative of the log of resistivity~\cite{Itoh2004PRL} can be also related to
the drastic changes at the characteristic temperature.
To confirm our proposal,
systematic measurements of $T$ dependence of X-ray diffraction
 as well as NMR 
 for single crystals are highly desired. 

Let us comment on the values of interchain Coulomb repulsions used in our calculations. 
By investigating several sets of parameters with varying $K^{-1}$, 
we have found that the mixed state is robust particularly for $V' \gg V''$, 
while the region becomes small when $V'$ is compatible to $V''$. 
The results imply that, 
since the center-to-center distances between two DI-DCNQI molecules are not
much different for $V'$ and $V''$, 
other factors are also important in determining 
the (effective) interchain Coulomb repulsions,  
such as the cation environment and the anisotropy of molecular orbitals.
In particular, the displacements of Ag$^+$ cations to positions
along the $V'$ bonds~\cite{Kakiuchi2007PRL} can be relevant by enhancing
the values of $V'$ (and reducing $V''$)
in our model, 
 since the cation with positive charge 
 enhances charge disproportionations between the neighboring DI-DCNQI molecules.
A similar effect was discussed in 
 another Q1D molecular conductor (TMTTF)$_2X$~\cite{Monceau2001PRL,Riera2001PRB}. 

Finally we make a remark on the spin degree of freedom. 
Experimentally, an antiferromagnetic transition is observed 
 by NMR~\cite{Hiraki1996PRB} and ESR~\cite{Sakurai2001JPSJ} at $T \simeq 5$ K. 
In the mixed state, 
we expect localized spins to appear 
 on the charge rich sites in the CO chains, 
 on the dimers in the LD chains, 
 and on the dimers in the coexisting chains 
 but with more spin density on the charge rich sites. 
Both intrachain and interchain exchange
 couplings depend on the degree of CO as well as LD, 
 which show characteristic $T$-dependences as discussed above; 
hence,  
it is highly nontrivial 
how the spin degree of freedom acts under such condition.
This should 
be a reason why simple Bonner-Fisher type analysis fails 
 to understand the magnetic susceptibility~\cite{Hiraki1996PRB} 
 and specific heat data~\cite{Nakazawa2002PRL}. 
Calculations including spins are left for future study.

In summary, 
 we have studied 
 a charge-frustrated spinless fermion model coupled to the lattice degree of freedom 
 to understand the charge-lattice coupled phenomena in 
the quasi-one-dimensional molecular conductor
 (DI-DCNQI)$_2$Ag. 
We have shown that 
a peculiar mixed state, where 
charge ordered, lattice dimerized, and their coexisting chains 
arrange periodically,  
 is self-organized so as to relieve the spiral frustration 
between
the chains.
We have proposed two scenarios for the finite-temperature properties, and
in both of them 
there is a characteristic temperature 
 where lattice distortions develop rapidly.
This drastic change 
provides a key to understand the puzzles remaining in experiments.
Similar interchain frustration is seen in other
quasi-one-dimensional compounds, and therefore, our results indicate
the necessity to examine its effect to obtain deeper understanding
of these systems. 

We acknowledge helpful discussions with 
S. Fujiyama, K. Hiraki, T. Itou, T. Kakiuchi, K. Kanoda, Y. Noda, and H. Sawa. 
This work is supported by the Grant-in-Aid for Scientific Research 
(Nos.~19014020, 19052008, 16GS0219) 
from MEXT. 

\footnotesize{$^{*}$Present address: Condensed Matter Theory Laboratory, RIKEN, Saitama 351-0198, Japan}


\begin{thebibliography}{99}
\bibitem{note0}
The order of $i$ and $j$ is set so that the transfer integrals and 
the Coulomb interactions increase 
when the intermolecular distance decreases.
\bibitem{note}
The relative values of $\beta$, $\beta'$, and $\beta''$ are set by assuming
each intersite Coulomb interaction to be $\propto 1/r_{ij}$ for simplicity
($r_{ij}$: inter-molecular distance).
\bibitem{note2}
In Fig.~\ref{fig3}(b), 
the CO+LD state always has lower energy 
near the dip verging on zero at $K^{-1}\sim0.23$.
\end{thebibliography}
\end{document}